\documentclass[
 reprint,
superscriptaddress,
 amsmath,amssymb,
aps,
prd]{revtex4-2}

\usepackage{graphicx}%
\usepackage{dcolumn}%
\usepackage{bm}%
\usepackage{xcolor}
\usepackage{physics}%
\usepackage{mathtools}%
\usepackage{hyperref}%
\usepackage{ulem}

\newcommand{\Xeff}{\chi_\text{eff}}
\newcommand{\Xp}{\chi_\text{p}}
\newcommand{\Mo}{M_\odot}

\usepackage{color}

\newcommand{\apjl}{The Astrophysical Journal Letters}
\newcommand{\aap}{A\&A}
\newcommand{\aj}{AJ}

\newcommand{\mnras}{MNRAS}
\DeclareMathOperator{\Dilog}{\mathrm{Li}_{2}}
\begin{document}

\preprint{APS/123-QED}

\title{
An analytical joint prior for effective spins\\ for inference on the spin distribution of binary black holes
}


\author{Masaki Iwaya}
 \email{iwaya@icrr.u-tokyo.ac.jp}
\affiliation{
 Institute for Cosmic Ray Research, University of Tokyo, Kashiwanoha 5-1-5, Kashiwa, Chiba 277-8582 Japan
}

\author{Kazuya Kobayashi}
\affiliation{
 Institute for Cosmic Ray Research, University of Tokyo, Kashiwanoha 5-1-5, Kashiwa, Chiba 277-8582 Japan
}

\author{Soichiro Morisaki}
\affiliation{
 Institute for Cosmic Ray Research, University of Tokyo, Kashiwanoha 5-1-5, Kashiwa, Chiba 277-8582 Japan
}

\author{Kenta Hotokezaka}
\affiliation{Research Center for the Early Universe, Graduate School of Science, University of Tokyo, 7-3-1 Hongo, Bunkyo-ku, Tokyo 113-0033, Japan}

\author{Tomoya Kinugawa}
\affiliation{Research Center for the Early Universe, Graduate School of Science, University of Tokyo, 7-3-1 Hongo, Bunkyo-ku, Tokyo 113-0033, Japan}
\affiliation{Faculty of Engineering, Shinshu University, 4-17-1, Wakasato, Nagano-shi, Nagano, 380-8553, Japan}
\affiliation{Research Center for Aerospace System, Shinshu University,  4-17-1, Wakasato, Nagano-shi, Nagano, 380-8553, Japan}

\date{\today}%

\begin{abstract}
We derive an analytical form of the joint prior of effective spin parameters, $\Xeff$ and $\Xp$, assuming an isotropic and uniform-in-magnitude spin distribution. This is a vital factor in performing hierarchical Bayesian inference for studying the population properties of merging compact binaries observed with gravitational waves. In previous analyses, this was evaluated numerically using kernel density estimation (KDE). However, we find that this numerical approach is inaccurate in certain parameter regions, where both $|\Xeff|$ and $\Xp$ are small. Our analytical approach provides accurate computations of the joint prior across the entire parameter space and enables more reliable population inference. Employing our analytic prior, we reanalyze binary black holes in the Gravitational-Wave Transient Catalog 3 (GWTC-3) by the LIGO-Virgo-KAGRA collaboration. While the results are largely unchanged, log-likelihood errors due to the use of the inaccurate prior evaluations are $\mathcal{O}(1)$. Since these errors accumulate with the increasing number of events, our analytical prior will be crucial in the future analyses.
\end{abstract}
\maketitle
\section{Introduction}
\label{sec: Intro}
About one hundred events have been detected \cite{GWTC-1, GWTC-2, GWTC-2.1, GWTC-3} since the first direct detection of gravitational waves (GWs) from the binary black hole (BBH) merger in 2015 \cite{GW150914}. The fourth observing run of the LIGO-Virgo-KAGRA collaboration (LVK) \cite{LIGO, VIRGO, KAGRA, LVK} is currently ongoing and expected to discover additional hundreds of BBH events. This growing number of events provides unprecedented findings about the nature of black holes. However, the astrophysical processes that govern the formation and evolution of BBHs remain largely unknown (see, e.g., \cite{Population_Review} for recent reviews).

Various BBH formation channels have been discussed, including isolated binaries, triples, and dynamical capture in globular clusters and Active Galactic Nuclei. These channels are further categorized based on specific characteristics, such as stellar metallicity and environmental conditions. Each formation channel exhibits a characteristic parameter distribution, which can be tested with BBHs observed by LVK  (e.g., \cite{Zevin2021, Iwaya2023}).
Using parametric models that extract the characteristics of these formation channels, LVK has 
estimated the distribution of source parameters (such as masses) and the associated merger rate density based on the latest GW transient catalogs \cite{GWTC-1_Pop, GWTC-2_Pop, GWTC-3_Pop}. In addition, non-parametric approaches have been used to capture the parameter distributions in a more model agnostic way (e.g. \cite{Nonpara1, Nonpara2}).

One promising approach for uncovering the origins of BBH systems is to measure black-hole (BH) spins \cite{Hotokezaka2017,farr2018using, Spin1, Spin2, Kinugawa2020,Tagawa2021MNRAS,Olejak2021,Samsing2022Natur,Trani2024A&A}. While individual spin components are difficult to measure, certain combinations of these components can be more reliably constrained. The effective inspiral spin, denoted by $\Xeff$, is one such parameter, defined by the mass-weighted average of the aligned spin components \cite{Xeff_OG}: 
\begin{equation}
\label{eq: Xeff}
    \Xeff\coloneqq\frac{a_1\cos\vartheta_1+qa_2\cos\vartheta_2}{1+q},
\end{equation}
where $a_i$ denotes the dimensionless spin magnitude of the $i$-th object, $\vartheta_i$ denotes the polar tilt angle, and $q \leq1$ is the mass ratio. Its distribution is a powerful tool to distinguish the origins of BBHs \cite{GerosaHierarchicalBH, Gerosa2018}.

Another important spin parameter is the effective precessing spin, denoted by $\Xp$, which  is  given by \cite{Xp0}
\begin{equation}
\label{eq: Xp}
\Xp\coloneqq \max\qty(a_1\sin\vartheta_1,\qty(\frac{3+4q}{4+3q})qa_2\sin\vartheta_2).
\end{equation} 
This parameter captures the magnitude of spin components perpendicular to the orbital angular momentum, which are responsible for precession effects in orbital motion.
Non-zero values of $\Xp$ indicate misalignments between the spins and the orbital angular momentum, which provides useful information to distingush the formation channels.
Some generalizations of this parameter have recently been discussed \cite{GeneralizedXp, HoyXp, HoyXp2020, Thomas}.

In the latest analysis by LVK with the Gravitational Wave Transient
Catalog 3 (GWTC-3) \cite{GWTC-3_Pop}, the distribution of $\Xeff$ and $\Xp$ is modeled as Gaussian distribution \cite{1stXeffGaussian}, and its mean and covariance are estimated.
According to their results,
the $\Xeff$ distribution is well described by a narrow Gaussian with a slightly positive mean $0.05$ and standard deviation around $0.1$.
For $\Xp$, the distribution is explained either by a flat distribution centered at $\Xp = 0$ or a narrow distribution centered around $\Xp\approx 0.2$.

Hierarchical Bayesian inference is commonly used to extract the population property of GW sources \cite{HBcite, HBcite2, HBcite3}.
The likelihood function in this inference involves the evidence for each event and selection function that characterizes observational selection effects.
Each component requires the evaluation of a high-dimensional integral over $\theta$.
These integrals are computed using Monte-Carlo methods, where random samples of $\theta$ are drawn from reference distributions, and Monte-Carlo sums are calculated with appropriate reweighting of them.
For evaluating the evidence, posterior samples are obtained from the parameter estimation of each event, and their weights in the reweighting process are inversely proportional to the prior distribution assumed in that analysis.
For evaluating the selection function, typically, a large number of simulated signals are injected into data to study the search sensitivities, and their weights are inversely proportional to the distribution from which the simulated signals are drawn.

For inferring the distribution of $\Xeff$ and $\Xp$, those reweighting processes require their joint distribution conditioned on $q$, denoted by $\pi(\Xeff, \Xp|q)$, computed from the prior distribution or the distribution populating the simulated signals.
Typically, these distributions are isotropic with respect to component spins and uniform in their magnitudes.
In the GWTC-3 analysis, it was evaluated numerically by drawing samples from the isotropic and uniform-in-magnitude spin distribution and applying kernel density estimation (KDE).

In this paper, we derive an analytical formula of $\pi(\Xeff,\Xp|q)$. 
By comparing our analytical method with the conventional numerical method, we find that the latter method fails to capture certain structures of $\pi(\Xeff,\Xp|q)$, introducing numerical errors in the population inference. Furthermore, this analytical approach is faster and eliminates the stochastic uncertainty inherent in the numerical methods.
To investigate potential biases in the GWTC-3 analysis, we reanalyzed the GWTC-3 BBHs with our analytic formula of $\pi(\Xeff,\Xp|q)$.

The structure of the paper is as follows: In Section \ref{sec: Priors}, we review the conventional numerical approach, then present an analytical approach, and evaluate how closely these approaches represent the true distribution. 
In section \ref{sec: HBI}, we re-analyze GWTC-3 BBH events with a model presented in \cite{GWTC-3_Pop} using numerical or analytical methods to calculate the prior distribution. Then we compare the results to quantitatively evaluate the bias caused by numerical methods. 
In section \ref{sec: Summary}, we summarize the findings and conclude the paper.

\section{Joint distribution of effective spins}
\label{sec: Priors}

In this section, we derive an analytic formula of $\pi(\Xeff, \Xp|q)$ and compare it with that computed with the conventional numerical method.
It is required to numerically compute the likelihood function of the population inference, which is given by
\begin{equation}
    \label{eq: Hyper-likelihood}
    p(\left\{ d_i \right\}|\Lambda) \propto \prod_{i=1}^{N_\mathrm{ev}}\frac{\mathcal{Z}_i(\Lambda)}{\alpha(\Lambda)}.
\end{equation}
$\mathcal{Z}_i(\Lambda)$ represents the evidence for the $i$--th event, given by
\begin{equation}
    \label{eq: event_evidence}
    \mathcal{Z}_i(\Lambda) = \int p_\mathrm{pop}(\theta_i|\Lambda) p(d_i|\theta_i)\dd{\theta_i},
\end{equation}
where $p_\mathrm{pop}(\theta|\Lambda)$ represents a parameterized model distribution of $\theta$, $\theta_i$ and $d_i$ represent the source parameters and data for the $i$--th event respectively, and $p(d_i|\theta_i)$ represents the likelihood function.
$\alpha(\Lambda)$ represents the selection function, given by
\begin{equation}
    \label{eq: alpha}
    \alpha(\Lambda) \coloneqq \int \dd\theta p_\mathrm{det}(\theta)p_\mathrm{pop}(\theta|\Lambda),
\end{equation}
where $p_\mathrm{det}(\theta)$ is the probability of detecting a GW signal with the source parameter values of $\theta$.

As explained in the previous section, those integrals are computed with Monte-Carlo methods.
$\mathcal{Z}_i(\Lambda)$ is computed with posterior samples from the parameter estimation of the $i$--th event, $\{\theta_i^j\}_{j=1}^{N^i_\mathrm{samp}}$:
\begin{equation}
    \label{eq: evidence_approx}
    \mathcal{Z}_i(\Lambda) \propto \frac{1}{N^i_\mathrm{samp}}\sum_{j=1}^{N^i_\mathrm{samp}}\frac{p_\mathrm{pop}(\theta_i^j|\Lambda)}{\pi_\mathrm{PE}(\theta_i^j)},
\end{equation}
where $\pi_\mathrm{PE}(\theta)$ is the prior distribution assumed in the parameter estimation.
$\alpha(\Lambda)$ is computed with simulated signals injected into data.
They are analyzed by search pipelines, and the Monte-Carlo sum is evaluated over the subset of signals successfully recovered by the pipelines:
\begin{equation}
    \label{eq: alpha_approx}
    \alpha(\Lambda) = \frac{1}{N_\mathrm{inj}}\sum_{k:\mathrm{found}}\frac{p_\mathrm{pop}(\theta_k|\Lambda)}{p_\mathrm{inj}(\theta_k)},
\end{equation}
where $N_\mathrm{inj}$ is the total number of injections, $p_\mathrm{inj}(\theta)$ is the probability density from which the injection source parameters are drawn, $\theta_k$ is the $k$--th recovered injection during the injection campaign.

In the LVK analyses, both $\pi_\mathrm{PE}(\theta)$ and $p_\mathrm{inj}(\theta)$ are isotropic in spin orientations and uniform in spin magnitudes.
Let $\bm{a}_i=(a_i, \cos\vartheta_i, \phi_i)$, where $\phi_i$ is the azimuthal angle of the $i$--th spin around the orbital angular momentum.
The isotropic and uniform-in-magnitude spin prior is given by
\begin{equation}
    \label{eq: OG_spin}
    \pi(\bm{a}_i)=\frac{1}{4\pi a_{\rm max}},
\end{equation}
where $a_{\rm max}$ denotes the maximum spin magnitude.
Here, we assume that the spin magnitudes of both compact objects share the same range, such that $0 \leq a_1, a_2 \leq a_{\rm max}$.

In population inference which focuses exclusively on $\Xeff$ and $\Xp$, the other spin components are marginalized over, resulting in the following expression for the joint prior:
\begin{align}
    \label{eq:本来の式}
    &\pi(\Xeff,\Xp|q)\notag\\
    =&\int\dd{\bm{a}_1}\dd{\bm{a}_2}
    \pi(\bm{a}_1)\pi(\bm{a}_2)\notag\\\times&
    \delta\qty(\Xeff-\frac{a_1\cos\vartheta_1+qa_2\cos\vartheta_2}{1+q})\notag\\\times&
    \delta\qty(\Xp-\max\qty(a_1\sin\vartheta_1,\qty(\frac{3+4q}{4+3q})qa_2\sin\vartheta_2)).
\end{align}
In the GWTC-3 analysis, this integral was numerically computed using KDE. 
We instead calculate this integral analytically to evaluate this joint prior more accurately.

\subsection{KDE prior}
\label{sec:KDEprior}
We begin by reviewing the numerical method employed in the GWTC-3 analysis. The joint distribution is decomposed as
\begin{equation}
    \pi(\Xeff,\Xp|q) = \pi(\Xp|\Xeff,q)\pi(\Xeff|q).
\end{equation}
The term $\pi(\Xeff|q)$ is evaluated with its analytical expression given by Eq. (10) in Ref. \cite{Xeffprior}, while $\pi(\Xp|\Xeff,q)$ is calculated numerically.

For the numerical evaluation of  $\pi(\Xp|\Xeff,q)$, KDE is applied to random samples drawn from the prior. Since this probability is conditioned by $\Xeff$, the actual degree of freedom for this draw is three \footnote{azimuthal angles are marginalized}. Samples of $a_1,a_2$, and $\cos\vartheta_2$ are drawn from the uniform priors, and the values of $\cos\vartheta_1$ are determined by Eq. \eqref{eq: Xeff} 
Unphysical samples with $\abs{\cos\vartheta_1}\geq1$ are rejected. This procedure is repeated until a sufficient number, 10,000 by default, of samples are obtained. Once a sufficient number of samples is obtained, $\Xp$ samples are generated according to \eqref{eq: Xp}, and KDE with the Gaussian kernel is applied to them. Due to the Jacobian of the transformation to $\Xeff$ from $\cos\vartheta_1$, each sample has the weight,
\begin{align}
    \pdv{\cos\vartheta_1(a_1,a_2,\Xeff,\cos\vartheta_2)}{\Xeff} = \frac{1+q}{a_1}.
\end{align}

The boundary conditions must be set carefully because  KDE, in general, tends to have biases near the boundaries \cite{Silverman}. In the numerical evaluation of $\pi(\Xp|\Xeff,q),$ 
the boundary condition is $\pi(0|\Xeff,q) = \pi(\chi_\mathrm{p,max}|\Xeff,q) = 0.$ Here, $\chi_\mathrm{p,max}$ is the physically maximum value for $\Xp$ given $\Xeff$ and $q$:
\begin{align}
\label{eq: Xpmax}
    &\chi_\mathrm{p,max}\coloneqq\max_{q,\Xeff}(\Xp)=\notag\\
    &\begin{dcases}
        a_\mathrm{max} & \frac{\abs{\Xeff}}{a_\mathrm{max}} \leq \frac{q}{1+q}\\
        \sqrt{a_\mathrm{max}^2-\qty((1+q)\abs{\Xeff}-a_\mathrm{max}q)^2}& \frac{\abs{\Xeff}}{a_\mathrm{max}} > \frac{q}{1+q}.
    \end{dcases}
\end{align}
To prevent the density from spilling out of $[0,\chi_\mathrm{p, max}]$, KDE for grids in $[\varepsilon\chi_\mathrm{p,max},(1-\varepsilon)\chi_\mathrm{p,max}]$ are evaluated first, where $\varepsilon$ is a small constant; $\varepsilon=0.02$ is chosen in the GWTC-3 inference. Then the boundary conditions are manually provided; other values are calculated by interpolation. Finally, the function is normalized to achieve
\begin{equation}
    \label{eq: normalization}
    \int_0^{\chi_\mathrm{p, max}}\pi(\Xp|\Xeff,q)=1.
\end{equation}
An implementation of this method can be found at \cite{GitHubCallister}, though with a different choice of $\varepsilon$.
$\pi(\Xp|\Xeff,q)$ as constructed above will henceforth be referred to as ``the KDE prior" in this paper.

This numerical method has several limitations. First, the use of KDE may obscure certain structures in the prior distribution.
For example, the KDE prior may not capture the discontinuous derivative behavior due to the max function in equation \eqref{eq:本来の式}.
Second, this method can be computationally expensive when dealing with a large number of events.
In particular, this method is slow if we give large $|\Xeff|$ because many random samples are rejected during the sampling process.
Third, this method introduces an additional source of uncertainty to the final results from the probabilistic nature of KDE. Because KDE relies on random samples, the results have statistical errors, but this is not an inherent property of the prior distribution.

\subsection{Analytical prior}
\label{sec: analyticalprior}
\begin{figure*}[htb]
\includegraphics[width=1.8\columnwidth]{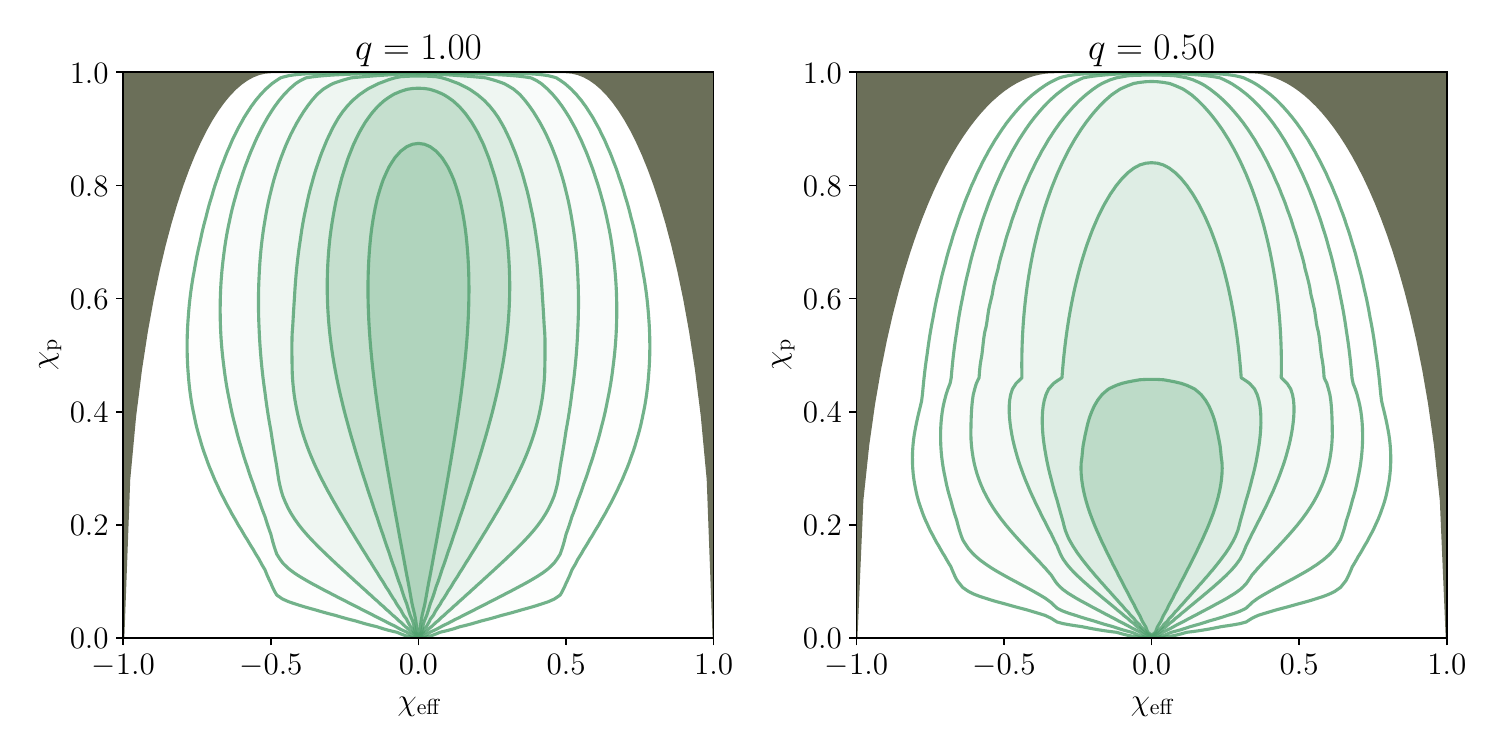}
\caption{\label{fig: Jointdist} The analytical joint prior on $\Xeff,\Xp$ for $q=1$ (left panel) and $q=0.5$ (right panel). Contours show $1/2\sigma$, $1\sigma$, \ldots, $3\sigma$ regions. In gray is shown the area where $\Xeff$--$\Xp$ pairs cannot physically be achieved simultaneously (obtained from \eqref{eq: Xpmax}).}
\end{figure*}

To overcome the limitations of the KDE prior, here we present an analytical joint prior.
The difficulty to compute equation \eqref{eq:本来の式} arises from the large number of conditional branches. In Ref. \cite{Xeffprior}, the author first specifies the branching cases and then provides primitive functions. However, the conditional expressions to be considered for computing $\pi(\Xeff,\Xp|q)$ are complex depending on $q,\Xeff,$ and $\Xp$. Furthermore, the number of cases to be considered is much larger than $\pi( \Xeff|q)$, and it is not easy to consider in advance how many cases will be needed.

Instead, we adopt an approach that we first identify the primitive function of the integral and then assign the appropriate integral domain. To achieve this procedure, we give the integral domain $a_i\in(0,1)$ and $\vartheta_i\in[0,\pi]$ by a step function:
\begin{equation}
    \Theta(x)=\begin{cases}
        1&x>0\\0&x<0.
    \end{cases}
\end{equation}

The following scaling relation with respect to $a_\mathrm{max}$ holds:
\begin{equation}
    \pi\left(\left.\Xeff,\Xp\right|q,a_\mathrm{max}=a\right)=\frac{1}{a^2}\pi\left(\left.\frac{\Xeff}{a},\frac{\Xp}{a}\right|q,a_\mathrm{max}=1\right).
\end{equation}
Thus, without loss of generality, we assume $a_\mathrm{max} = 1$ hereafter.
By the simple transformation of the integral variables $z_i\coloneqq a_i\cos\vartheta_i$ and $x_i\coloneqq a_i\sin\vartheta_i$, we get
\begin{widetext}
\begin{align}
    \pi\left(\left.\Xeff,\Xp\right|q\right)
    &=
    \frac{1}{4}
    \int\dd{z_1}
    \int\dd{z_2}
    \int\dd{x_1}
    \int\dd{x_2}
    \frac{x_1}{x_1^2+z_1^2}
    \frac{x_2}{x_2^2+z_2^2}
    \Theta(x_1)
    \Theta(x_2)
    \notag\\&\times
    \delta\qty(\Xeff - \frac{z_1+qz_2}{1+q})
    \delta\qty(\Xp-\max\qty(x_1,\frac{3+4q}{4+3q}qx_2))
    \Theta\qty(1-x_1^2-z_1^2)
    \Theta\qty(1-x_2^2-z_2^2)
    .
\end{align}
\end{widetext}

The integral can be reduced to the sum of one-dimensional integrals as (see appendix \ref{sec: Derivation} for details)
\begin{equation}
    \pi(\Xeff,\Xp|q)=I_1+I_2+I_3+I_4,
\end{equation}
where
\begin{widetext}
    \begin{align}
    \label{eq: I1}
    I_1 &= \frac{1+q}{8q}\Theta(\Xp)\Theta\qty(q-\frac{4+3q}{3+4q}\Xp)\Theta(x_1^\mathrm{max}-x_1^\mathrm{min})\notag\\
    &\times\qty[F\qty(x_1^\mathrm{max}\left|(1+q)\Xeff,\Xp,\frac{4+3q}{3+4q}\Xp,q\right.)-F\qty(x_1^\mathrm{min}\left|(1+q)\Xeff,\Xp,\frac{4+3q}{3+4q}\Xp,q\right.)]\\
    \label{eq: I2}
    I_2 &= -
    \frac{1+q}{8q}\Theta(\Xp)\Theta\qty(1-\Xp)\Theta(x_2^\mathrm{max}-x_2^\mathrm{min})\qty[F\qty(x_2^\mathrm{max}\left|(1+q)\Xeff,\Xp,0,q\right.)-F\qty(x_2^\mathrm{min}\left|(1+q)\Xeff,\Xp,0,q\right.)]\\
    \label{eq: I3}
    I_3 &= \frac{1+q}{8q}\frac{4+3q}{3+4q}\Theta(\Xp)\Theta\qty(q-\frac{4+3q}{3+4q}\Xp)\Theta(x_3^\mathrm{max}-x_3^\mathrm{min})\notag\\&\times\qty[F\qty(x_3^\mathrm{max}\left|(1+q)\Xeff,\frac{4+3q}{3+4q}\Xp,\Xp,1\right.)-F\qty(x_3^\mathrm{min}\left|(1+q)\Xeff,\frac{4+3q}{3+4q}\Xp,\Xp,1\right.)]\\
    \label{eq: I4}
    I_4 &= -\frac{1+q}{8q}\frac{4+3q}{3+4q}\Theta(\Xp)\Theta\qty(q-\frac{4+3q}{3+4q}\Xp)\Theta(x^\mathrm{max}_4-x^\mathrm{min}_4)\notag\\&\times\qty[F\qty(x^\mathrm{max}_4\left|(1+q)\Xeff,\frac{4+3q}{3+4q}\Xp,0,1\right.)-F\qty(x^\mathrm{min}_4\left|(1+q)\Xeff,\frac{4+3q}{3+4q}\Xp,0,1\right.)]
\end{align}
and
\begin{align}
    \label{eq: x1}
    &\begin{dcases}
    x_1^\mathrm{max}=\min\qty(\sqrt{q^2-\qty(\frac{4+3q}{3+4q})^2\Xp^2},(1+q)\Xeff+\sqrt{1-\Xp^2})\\
    x_1^\mathrm{min}=\max\qty(-\sqrt{q^2-\qty(\frac{4+3q}{3+4q})^2\Xp^2},(1+q)\Xeff-\sqrt{1-\Xp^2})
    \end{dcases}\\
    \label{eq: x2}
    &\begin{dcases}
    x_2^\mathrm{max}=\min\qty(q,(1+q)\Xeff+\sqrt{1-\Xp^2})\\
    x_2^\mathrm{min}=\max\qty(-q,(1+q)\Xeff-\sqrt{1-\Xp^2})
    \end{dcases}\\
    \label{eq: x3}
    &\begin{dcases}
    x_3^\mathrm{max}=\min\qty(\sqrt{1-\Xp^2},(1+q)\Xeff+\sqrt{q^2-\qty(\frac{4+3q}{3+4q})^2\Xp^2})\\
    x_3^\mathrm{min}=\max\qty(-\sqrt{1-\Xp^2},(1+q)\Xeff-\sqrt{q^2-\qty(\frac{4+3q}{3+4q})^2\Xp^2})
    \end{dcases}\\
    \label{eq: x4}
    &\begin{dcases}
    x_4^\mathrm{max}=\min\qty(1,(1+q)\Xeff+\sqrt{q^2-\qty(\frac{4+3q}{3+4q})^2\Xp^2})\\
    x_4^\mathrm{min}=\max\qty(-1,(1+q)\Xeff-\sqrt{q^2-\qty(\frac{4+3q}{3+4q})^2\Xp^2}).
    \end{dcases}
\end{align}
\end{widetext}
These expressions have roots whose contents can be negative, but they are greater than or equal to zero because of step functions.
The one-dimensional function $F(x|a,b,c,d)$ is defined as
\begin{equation}
    F(x|a,b,c,d)\coloneqq\int_0^x \dd{x}\frac{b}{(x-a)^2+b^2}\log(\frac{x^2+c^2}{d^2}),
\end{equation}
and has an analytical form:
\begin{widetext}
    \begin{align}
        \label{eq: F}
        &F(x|a,b,c,d)=G\left(\left.\frac{x}{b}\right|\frac{a}{b},\frac{c}{b}\right)+\log(\frac{b^2}{d^2})\qty[\arctan(\frac{x-a}{b})+\arctan(\frac{a}{b})],\\
        \label{eq:G}
        &G(x|\alpha,\beta) =
        \mathrm{Im} [g(x|\alpha, \beta) + g(x|\alpha, -\beta) - g(0|\alpha, \beta) - g(0|\alpha, -\beta)],\\
    \label{eq:g}
    &g(x|\alpha,\beta)
    =
    \begin{dcases}
        \log(x-\beta i)\log(\frac{\alpha -x+i}{\alpha +i- \beta i})+\Dilog\qty(\frac{x-\beta i}{\alpha +i-\beta i})& |\beta|<1\\
        \frac{1}{2}\qty(\log(x-\alpha-i))^2+\Dilog\qty(\frac{-\alpha}{x-\alpha-i})&\beta =1,\alpha\leq0\\
        \log(\alpha+i-\beta i)\log(\alpha-x+i)-\Dilog\qty(\frac{\alpha-x+i}{\alpha+i-\beta i})& \mathrm{otherwise},\\
    \end{dcases}\\
    \label{eq:Dilog}
    &\Dilog(z)=-\int_0^z\dd{z'}\frac{\log(1-z')}{z'}.
\end{align}
\end{widetext}
The branching of $g(x|\alpha,\beta)$ arises in order to avoid transcending branch cuts of complex logarithm function and the dilogarithm function denoted by $\Dilog(z)$, which are both on the real axis. The more detailed derivation of an analytical form of $F(x|a,b,c,d)$ is given in Appendix \ref{sec: Derivation}. These expressions provide the fully analytical formulae to compute $\pi(\Xeff,\Xp|q).$ We will refer to this approach as ``the analytical prior''.

The analytical prior in two-dimensional plane is shown in Figure \ref{fig: Jointdist}. In both cases, the joint distributions are symmetric around $\Xeff=0$. The two-dimensional distribution takes its supremum at the limit of $\Xeff=0,\Xp\to0$.
The larger the value of $\left|\Xeff\right|$ is, the slower the rise of $\pi(\Xp|\Xeff,q)$ at $\Xp=0$ is. The contours are nearly parallel to the $\Xp=\mathrm{const.}$ line near the $\Xp=1$ limit, suggesting the steep decline at the $\Xp=1$ boundary. In addition, the right panel ($q=0.5$) illustrates that the contours have cusps around $\Xp\approx0.4$, which is resulted from the behavior of the $\max$ function.

\subsection{Comparisons with the KDE prior}
\label{sec:vsKDEprior}
\begin{figure*}[htb]
\includegraphics[width=1.8\columnwidth]{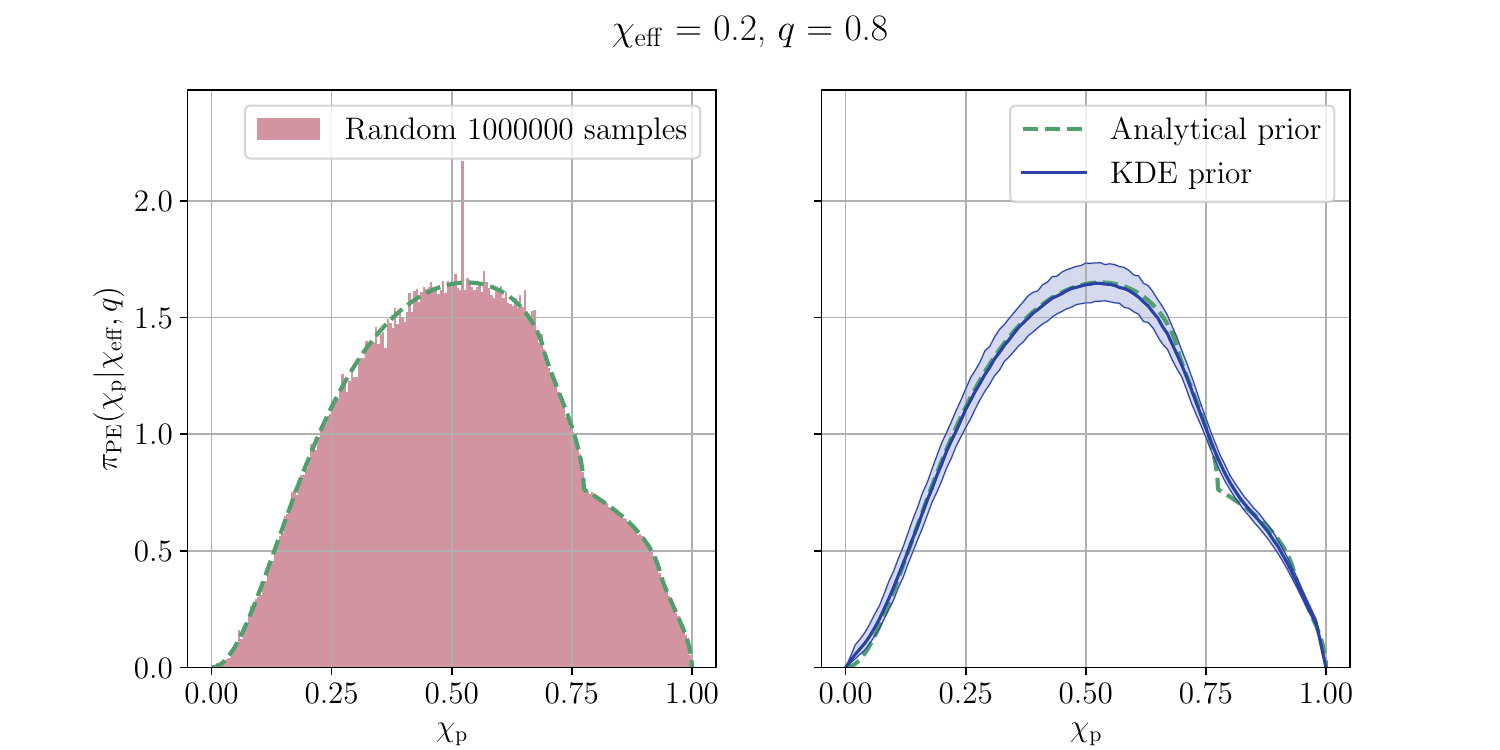}
    \caption{Comparison of different evaluation methods for $\pi(\Xp|\Xeff,q)$. For both panels, the analytical evaluation of $\pi(\Xp|\Xeff,q)$ is shown in green. In the left panel, the red histogram shows the behavior of random samples drawn from uniform-in-magnitude and isotropic spin prior. In the right panel, in blue is the evaluation of $\pi(\Xp|\Xeff,q)$ by the KDE prior. The blue bold line indicates the median of the evaluation and the blue colored area indicates the 90\% region. Given parameters are $\Xeff=0.2,q=0.8$.}
    \label{fig: Xp_given_Xeff_q_1}
\end{figure*}
\begin{figure*}[htb]
\includegraphics[width=1.8\columnwidth]{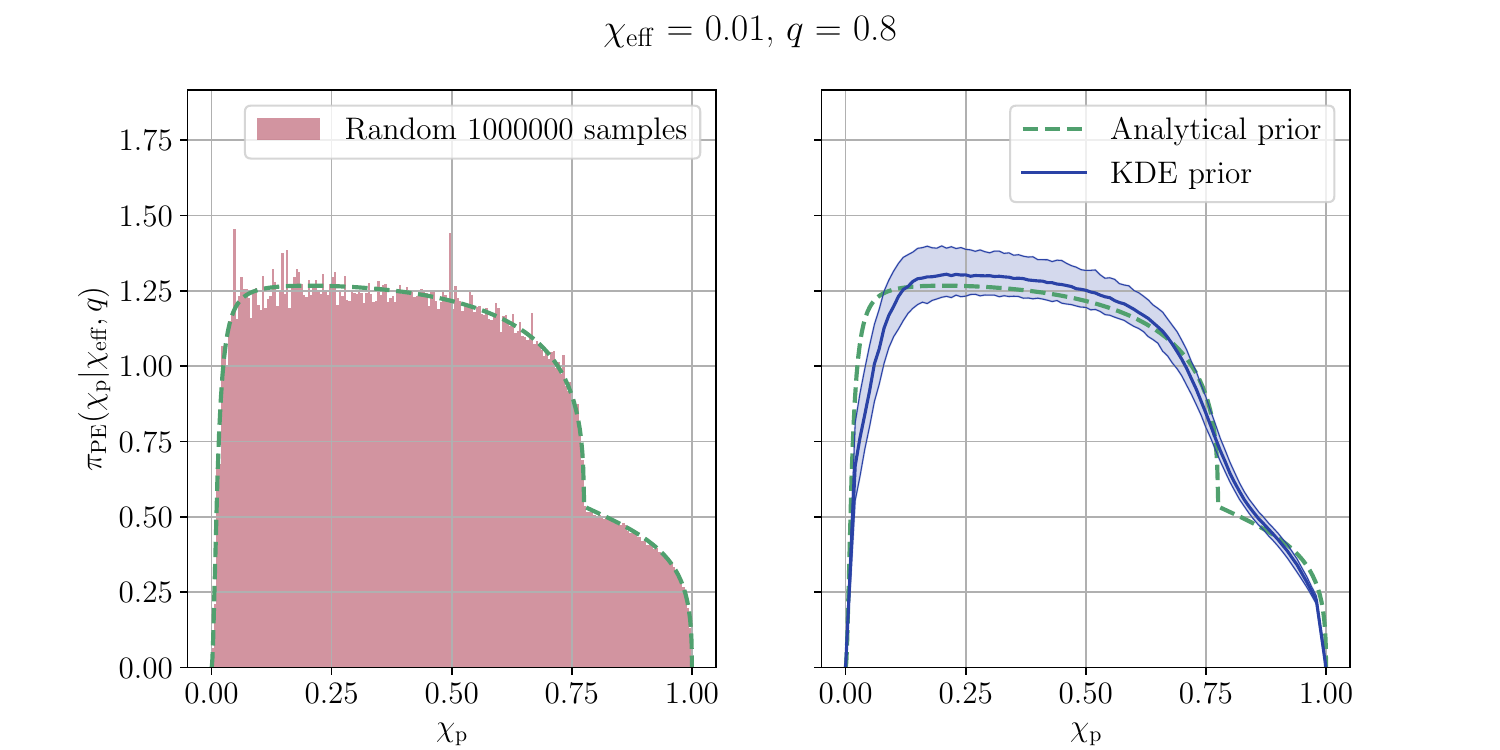}
    \caption{Same as Fig. \ref{fig: Xp_given_Xeff_q_1}, but we give $\Xeff=0.01,q=0.8$.}
    \label{fig: Xp_given_Xeff_q_2}
\end{figure*}

Figures \ref{fig: Xp_given_Xeff_q_1} and \ref{fig: Xp_given_Xeff_q_2} compare 
the analytical prior with histograms from random samples or with the KDE prior, in terms of a one-dimensional distribution $\pi(\Xp|\Xeff,q)$. 
The KDE prior is evaluated 1000 times. Its median and 90\% interval are shown to capture the probabilistic uncertainty. 

Figure \ref{fig: Xp_given_Xeff_q_1} shows the case of $\Xeff=0.2,q=0.8$. The analytical prior and the histogram are in good agreement. In addition, the analytical prior and the KDE prior are in good agreement except for around $\Xp=0.775$, where the analytical prior has a cusp.
This cusp corresponds to the point where the second input to the max function, $((3+4q)/(4+3q)) qa_2\sin\vartheta_2$, becomes physically impossible to be $\Xp$:
\begin{align}
    &\Xp =\notag\\ &\begin{dcases}
        \frac{3+4q}{4+3q}q&(1+q)\abs{\Xeff}\leq1\\
        \frac{3+4q}{4+3q}\sqrt{q^2-\qty(1-(1+q)\abs{\Xeff})^2}&(1+q)\abs{\Xeff}\geq1,
    \end{dcases}
\end{align}
and appears as the point where $I_4$ becomes zero because of the step functions.

Figure \ref{fig: Xp_given_Xeff_q_2} depicts the case of $\Xeff=0.01,q=0.8$. We see notable differences between the analytical and the KDE priors around the boundaries and the cusp. In addition, the median of the KDE prior is larger than the analytical prior for $0.2 \lesssim \Xp \lesssim 0.7$. The bandwidth of the KDE, constructed from 10000 random samples, is around $0.2$. This value is too large to effectively capture the sharp behaviors near the boundaries and the cusp.
This results not only in discrepancies around the boundaries and cusp, but also in the overestimation in $0.2 \lesssim \Xp \lesssim 0.7$ through the normalization procedure \eqref{eq: normalization}.
This bias can be reduced by increasing the number of samples, though this comes at the expense of higher computational costs.
Such discrepancies are generally unavoidable for small values of $|\Xeff|$ and can have a substantial impact on population inference when the underlying distribution is concentrated in that parameter region.

\section{Re-analyses of GWTC-3 BBH population}
In this section, we reanalyze GWTC-3 BBHs
using both the analytical prior and the KDE prior, then compare their results.
\label{sec: HBI}
\subsection{Model description}

\label{sec:ModelDescription}
In the GWTC-3 population analysis, LVK studied the spin distribution with the \textsc{Gaussian spin model} \cite{1stXeffGaussian, GWTC-3_Pop}.
The source parameters of interest in this model are $(m_1,q,z,\Xeff,\Xp)$ and the distributions of these parameters are described by the parametric functions:
\begin{widetext}
\begin{align}
    p_\mathrm{pop}(m_1|\Lambda) &= f_\mathrm{peak}\qty[\frac{1}{\sqrt{2\pi\sigma_m^2}}\exp(-\frac{1}{2}\qty(\frac{m_1-\mu_m}{\sigma_m})^2)]+(1-f_\mathrm{peak})\frac{(1+\alpha)m_1^\alpha}{(M_\mathrm{Max}^{1+\alpha} - M_\mathrm{Min}^{1+\alpha})},\\
    p_\mathrm{pop}(q|\Lambda) &\propto q^{\beta_q},\\
    \label{eq:zmodel}
    p_\mathrm{pop}(z|\Lambda) &\propto (1+z)^{\kappa-1}\dv{V_\mathrm{c}}{z},\\
    \label{eq: Ppop_spins}
    p_{\rm pop}(\Xeff,\Xp|\Lambda)&\propto 
    \exp[
    -\frac{1}{2(1-\rho^2)}
    \qty[
    \qty(\frac{\Xeff-\mu_{\rm eff}}{\sigma_{\rm eff}})^2
    -2\rho
    \qty(\frac{\Xeff-\mu_{\rm eff}}{\sigma_{\rm eff}})
    \qty(\frac{\Xp-\mu_{\rm p}}{\sigma_{\rm p}})
    +\qty(\frac{\Xp-\mu_{\rm p}}{\sigma_{\rm p}})^2
    ]
    ].
\end{align}
\end{widetext}
The hyper-parameters included in $\Lambda$ and their prior are summarized in Tab. \ref{tab: HyperPriorSummary}.

We analyze the same set of the BBH events analyzed in the GWTC-3 population inference for the \textsc{Gaussian spin model}, which comprises 69 BBHs in total.
For evaluating $\mathcal{Z}_i$, we use the posterior samples available at \cite{GWOSC}.
We use the same sets of posterior samples employed in the GWTC-3 population inference: \texttt{Overall\_posterior} samples for O1/O2 events, \texttt{PrecessingIMRPHM} samples for the new events reported in GWTC-2, and \texttt{C01:Mixed} samples for the new events in GWTC-3.
We use the so-called ``Nocosmo" samples, which are the posterior samples obtained with the distance prior $\pi_\mathrm{PE}(D_\mathrm{L})\propto D_\mathrm{L}^2$.
For evaluating $\alpha(\Lambda)$, we adopt a set of public injections \cite{Injections}. 
We also employ the same inference code used for the GWTC-3 analysis.
It is available at \cite{GitLigoCallister} and built upon the \texttt{emcee} package \cite{emcee}.
For the calculation of redshift distribution \eqref{eq:zmodel}, the Planck15 cosmology \cite{Planck} implemented in \texttt{astropy} \cite{astropy:2013,astropy:2018,astropy:2022} is adopted:
\begin{equation}
    H_0=67.9\ \mathrm{km\ s^{-1}\ Mpc^{-1}}, \Omega_\mathrm{m}=0.3065.
\end{equation}

Approximating integrals using Monte Carlo sums can yield unphysical results when a small number of samples dominate the total contribution. 
To avoid such situations, the inference has thresholds based on the effective sample size of the Monte-Carlo integration, which is defined by:
\begin{equation}
    N_\mathrm{eff}=\frac{\qty(\sum_i w_i)^2}{ \sum_i w_i^2},
\end{equation}
where $w_i$ is the weight for the $i$-th sample \cite{Kish}. 
We impose the same threshold as in the GWTC-3 analysis:
$N_\mathrm{eff} > 10$ for evaluating $\mathcal{Z}_i$, and $N_\mathrm{eff} > 4N_\mathrm{ev}$ for $\alpha(\Lambda)$ \cite{SelectionEffect}. If these conditions are not satisfied, $p(\{d_i\}|\Lambda)$ is set to zero.
In the GWTC-3 population inference \cite{GWTC-3_Pop}, the number of posterior samples used for evaluating $\mathcal{Z}_i$ was reduced to 4000, but we found that this may impose a significant limitation on the region of prior space due to the $N_\mathrm{eff}$ threshold. Therefore, our analysis uses all the posterior samples per event. We confirm that if the sample size is reduced to 4000, the result of the inference with the KDE prior is consistent with the result shown in \cite{GWTC-3_Pop}.

 \begin{table*}[htb]
\caption{\label{tab: HyperPriorSummary}%
Summary of hyper-parameters we investigate. In the last column, $U(a,b)$ means uniform distribution between $a$ and $b$, while $G(\mu,\sigma)$ means Gaussian distribution whose mean is $\mu$ and standard deviation is $\sigma$. For $M_\mathrm{min}$, we fix the value to be $5M_\odot$.
}
\begin{ruledtabular}
\begin{tabular}{lccc}
\textrm{Hyper-parameter}&
\textrm{Related source parameter}&
\textrm{Description}&
\textrm{Hyper-prior}\\
\colrule
$\alpha$&$m_1$&Power-law index for $m_1$ distribution&$U(-5,4)$\\
$M_\mathrm{max}$&$m_1$& Max value for power-law part of $m_1$&$U(60 \Mo,100 \Mo)$\\
$M_\mathrm{min}$&$m_1$& Min value for power-law part of $m_1$&$5 \Mo$\\
$\mu_m$&$m_1$& Gaussian mean for peak part of $m_1$&$U(20\Mo,100\Mo)$\\
$\sigma_m$&$m_1$& Gaussian width for peak part of $m_1$&$U(1\Mo,10\Mo)$\\
$f_\mathrm{peak}$&$m_1$& Fraction of masses in Gaussian part in $m_1$ distribution &$U(0,1)$\\
$\beta_q$&$q$& Power-law index for $q$ distribution &$U(-2,10)$\\
$\kappa$&$z$& Power-law index for $z$ distribution &$G(\mu=0,\sigma=6)$\\
$\mu_\mathrm{eff}$&$\Xeff$& Gaussian mean for $\Xeff$ distribution &$U(-1,1)$\\
$\sigma_\mathrm{eff}$&$\Xeff$& Gaussian width for $\Xeff$ distribution &$U(0.05,1)$\\
$\mu_\mathrm{p}$&$\Xp$& Gaussian mean for $\Xp$ distribution&$U(0.05,1)$\\
$\sigma_\mathrm{p}$&$\Xp$& Gaussian width for $\Xp$ distribution&$U(0.05,1)$\\
$\rho$&$\Xeff,\Xp$& Correlation between $\Xeff$ and $\Xp$ &$U(-0.75,0.75)$\\
\end{tabular}
\end{ruledtabular}
\end{table*}

\subsection{Results of the inference}

\label{sec: Inferenceresult}
\begin{widetext}
\begin{figure*}[htb]
\includegraphics[width=1.5\columnwidth]{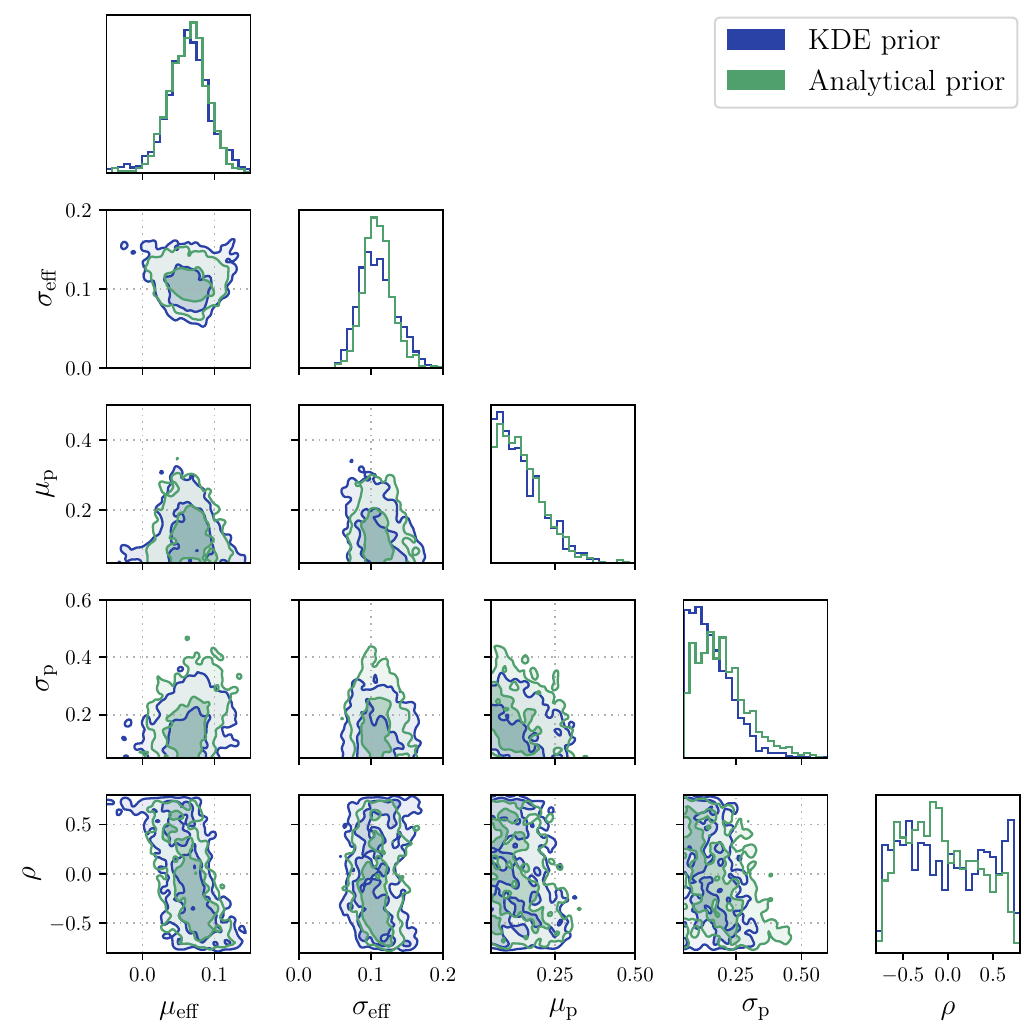}
\caption{Posterior distribution for hyper parameters that impact the $\Xeff$-$\Xp$ distributions.}
\label{fig: Corner_spins}
\end{figure*}    
\end{widetext}

\begin{figure*}[htb]
\includegraphics[width=2\columnwidth]{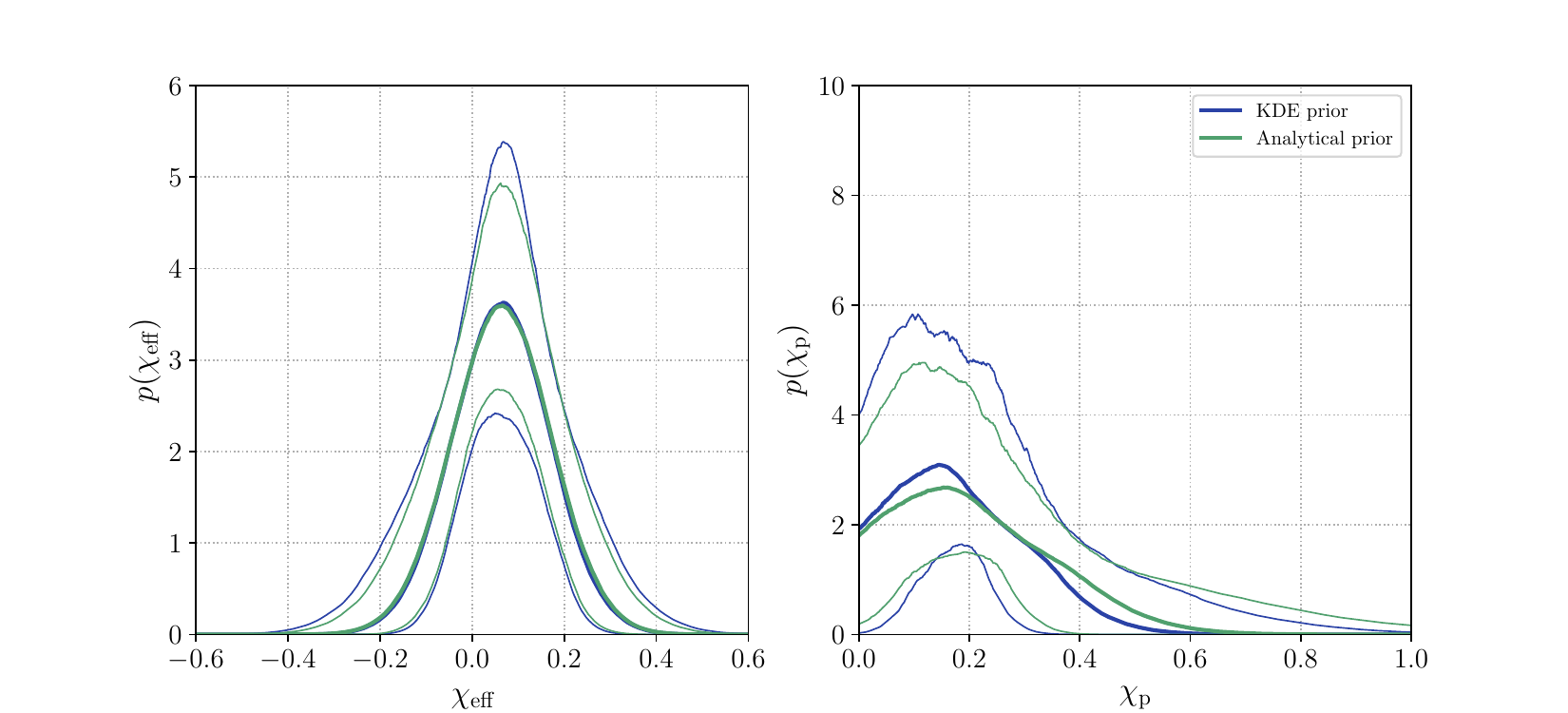}
\caption{
The recovered distributions for $\Xeff$ (left panel) and $\Xp$ (right panel).
The thick lines show the median of the recovered distribution, while the thinner lines show the 90\% intervals.
} 
\label{fig: PPD}
\end{figure*}

Figure \ref{fig: Corner_spins} shows the posterior distributions of hyper parameters related to $\Xeff$ and $\Xp$. These distributions remain largely unaffected by the prior replacement, although small differences are observed in $\sigma_\mathrm{eff}$ and $\sigma_\mathrm{p}$. The distribution of $\sigma_\mathrm{eff}$ gets narrower, with its $90\%$ credible interval changing from $\sigma_\mathrm{eff}=0.108^{+0.047}_{-0.035}$ to $\sigma_\mathrm{eff}=0.109^{+0.036}_{-0.030}$. The distribution of $\sigma_{\mathrm{p}}$ shifts to higher values, with its $90\%$ credible interval changing from $\sigma_\mathrm{p}=0.142^{+0.163}_{-0.085}$ to $\sigma_\mathrm{p}=0.176^{+0.149}_{-0.105}$.

Figure \ref{fig: PPD} shows the recovered distributions for $\Xeff$ and $\Xp$. Reflecting the subtle changes in $\sigma_\mathrm{eff}$ and $\sigma_\mathrm{p}$, the distribution of $\Xeff$ becomes slightly narrower, while the distribution of $\Xp$ broadens slightly. No notable differences are observed for hyper parameters related to masses and redshift.

\begin{figure}[htb]
\includegraphics[width=0.99\columnwidth]{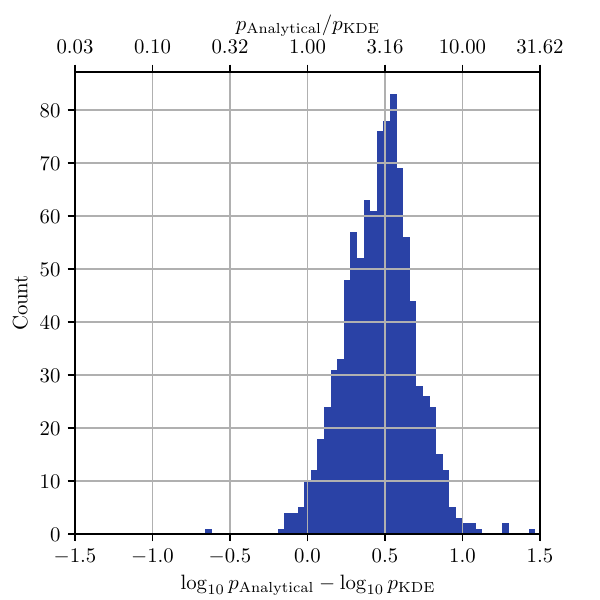}
\caption{The differences between the values of $\log_{10} p(\{d_i\}|\Lambda)$ computed with the analytical and KDE priors.} 
\label{fig: log_difference}
\end{figure}

Figure \ref{fig: log_difference} shows the errors in $\log_{10} p(\{d_i\}|\Lambda)$ due to the use of the KDE prior, computed on the samples of $\Lambda$ obtained by the inference using the analytical prior.
They are computed as the differences between the values of $\log_{10} p({d_i}|\Lambda)$ computed with the analytical and KDE priors.
Around $15\%$ of the samples do not meet the thresholds for $N_\mathrm{eff}$, and errors are not computed for these samples.
As shown in the figure, the errors are already non-negligible and can reach $\sim 1$.

The errors are expected to grow as the number of events increases. 
It can be understood through the following relationship between $\log_{10} p(\{d_i\}|\Lambda)$, $\log_{10} \alpha(\Lambda)$, and $\log_{10} \mathcal{Z}_i$:
\begin{equation}
    \log_{10} p(\{d_i\}|\Lambda) = N_\mathrm{ev} \log_{10} \alpha(\Lambda) + \sum_{i=1}^{N_{\mathrm{ev}}} \log_{10} \mathcal{Z}_i + \mathrm{const.} 
\end{equation}
The use of the KDE prior introduces small errors in $\log_{10} \alpha(\Lambda)$ and $\log_{10} \mathcal{Z}_i$, which accumulate according to this equation as $N_{\mathrm{ev}}$ increases.
Therefore, this issue is likely to become critical in future analyses, such as the analyses of new events detected in the LVK's fourth observing run (O4).
Consequently, our analytical prior will play an essential role.


\section{Conclusions}
\label{sec: Summary}
We derive an analytical expression for the joint prior distribution of effective spin parameters $\Xeff$ and $\Xp$ given isotropic and uniform-in-magnitude spins. 
By comparing our analytical prior with the KDE prior, computed with the numerical approach employed in Refs \cite{GWTC-2_Pop, GWTC-3_Pop}, we found that the latter approach causes inaccurate assessments of $\pi(\Xeff,\Xp|q)$ in a certain parameter region, especially in the area of $\abs{\Xeff}\lesssim0.01$. With the analytical prior, we re-analyze the GWTC-3 BBHs reported by LVK and estimate their spin distribution by adopting the \textsc{Gaussian Spin Model} employed in the GWTC-3 analysis. The results are largely unchanged from those obtained with the KDE prior.
However, the errors in likelihood caused by using the KDE prior are non-negligible. Since they are expected to grow as the number of events increases, our analytical prior will play an essential role in future analyses incorporating new GW events.
While this paper only examines and discusses the application of the analytical joint prior to the \textsc{Gaussian Spin Model}, this prior distribution is widely required in hierarchical inference for models that incorporate the two spin parameters $\Xeff$ and $\Xp$.

\section*{Acknowledgement}
The authors thank Thomas Callister for the development of the original code of GWTC-3 \textsc{Gaussian Spin Model} and for sharing his code. We would like to thank Daiki Watari, Hayato Imafuku, and Takahiro S Yamamoto for fruitful discussions. M. I. is supported by Forefront Physics and Mathematics Program to Drive Transformation (FoPM), a World-leading Innovative Graduate Study (WINGS) Program, the University of Tokyo. S. M. acknowledges support from JSPS Grant-in-Aid for Transformative Research Areas (A) No.~23H04891 and No.~23H04893. T. K. is supported from JSPS Grant-in-Aid for Transformative Research Areas (C) No.~22K03630.
K. H. work was supported by JSPS KAKENHI Grant-in-Aid for Scientific Research JP23H01169, JP23H04900, and JST FOREST Program JPMJFR2136.

This research has made use of data or software obtained from the Gravitational Wave Open Science Center (gwosc.org), a service of the LIGO Scientific Collaboration, the Virgo Collaboration, and KAGRA. This material is based upon work supported by NSF's LIGO Laboratory which is a major facility fully funded by the National Science Foundation, as well as the Science and Technology Facilities Council (STFC) of the United Kingdom, the Max-Planck-Society (MPS), and the State of Niedersachsen/Germany for support of the construction of Advanced LIGO and construction and operation of the GEO600 detector. Additional support for Advanced LIGO was provided by the Australian Research Council. Virgo is funded, through the European Gravitational Observatory (EGO), by the French Centre National de Recherche Scientifique (CNRS), the Italian Istituto Nazionale di Fisica Nucleare (INFN) and the Dutch Nikhef, with contributions by institutions from Belgium, Germany, Greece, Hungary, Ireland, Japan, Monaco, Poland, Portugal, Spain. KAGRA is supported by Ministry of Education, Culture, Sports, Science and Technology (MEXT), Japan Society for the Promotion of Science (JSPS) in Japan; National Research Foundation (NRF) and Ministry of Science and ICT (MSIT) in Korea; Academia Sinica (AS) and National Science and Technology Council (NSTC) in Taiwan.

\appendix

\begin{widetext}
\section{Detailed derivation of the analytical joint prior}
\label{sec: Derivation}
Our goal is to provide the analytical form of \eqref{eq:本来の式}. It is easy to integrate with azimuthal angles, as both $\Xeff$ and $\Xp$ do not depend on these parameters. With the definitions of $x_i = a_i\sin\vartheta_i, z_i = a_i\cos\vartheta_i$ we have
\begin{align}
    p(\Xeff,\Xp|q,a_{\rm max})
    &=
    \label{Def}
    \frac{1}{4a_{\rm max}^2}
    \int\dd{z_1}
    \int\dd{z_2}
    \int\dd{x_1}
    \int\dd{x_2}
    \frac{x_1}{x_1^2+z_1^2}
    \frac{x_2}{x_2^2+z_2^2}
    \notag\\&\times
    \delta\qty(\Xeff - \frac{z_1+qz_2}{1+q})
    \delta\qty(\Xp-\max\qty(x_1,\frac{3+4q}{4+3q}qx_2))
    \notag\\&\times
    \Theta(x_1)
    \Theta(x_2)
    \Theta\qty(a_{\rm max}^2-x_1^2-z_1^2)
    \Theta\qty(a_{\rm max}^2-x_2^2-z_2^2)
    .
\end{align}
By scaling all the integral variables like $x_1 = a_{\rm max}x_1',$ it can be shown that
\begin{equation}
    \pi\left(\left.\Xeff,\Xp\right|q,a_\mathrm{max}=a\right)=\frac{1}{a^2}\pi\left(\left.\frac{\Xeff}{a},\frac{\Xp}{a}\right|q,a_\mathrm{max}=1\right),
\end{equation}
so we assume $a_\mathrm{max} = 1$ and will not refer to it in this appendix. The following equation
\begin{equation}
    1 = \Theta\qty(x_1- \frac{3+4q}{4+3q} qx_2) + \Theta\qty(\frac{3+4q}{4+3q} qx_2 -x_1)
\end{equation}
can be used to divide the integral into two parts that provide different results for the max function. Then we can perform the integral with respect to both $x_1$ and $x_2$. These give
    \begin{align}
    \pi(\Xeff,\Xp|q)
    &=
    \frac{1}{8}
    \int\dd{z_1}
    \int\dd{z_2}
    \frac{\Xp}{\Xp^2+z_1^2}
    \log(z_2^2+\qty(\frac{4+3q}{3+4q})^2\frac{\Xp^2}{q^2})
    \delta\qty(\frac{z_1+qz_2}{1+q}-\Xeff)
    \notag\\&\times
    \Theta(\Xp)
    \Theta\qty(1-\Xp^2-z_1^2)
    \Theta\qty(q^2-q^2z_2^2-\qty(\frac{4+3q}{3+4q})^2\Xp^2)
    \notag\\
    &-
    \frac{1}{8}
    \int\dd{z_1}
    \int\dd{z_2}
    \frac{\Xp}{\Xp^2+z_1^2}
    \log(z_2^2)
    \delta\qty(\frac{z_1+qz_2}{1+q}-\Xeff)
    \notag\\&\times
    \Theta(\Xp)
    \Theta\qty(1-\Xp^2-z_1^2)
    \Theta\qty(1-z_2^2)
    \notag\\
    &+
    \frac{1}{8}\frac{4+3q}{3+4q}
    \int\dd{z_1}
    \int\dd{z_2}
    \frac{\qty(\frac{4+3q}{3+4q})\Xp}{\qty(\frac{4+3q}{3+4q})^2\Xp^2+z_2^2}
    \log(z_1^2+\Xp^2)
    \delta\qty(\frac{z_1+qz_2}{1+q}-\Xeff)
    \notag\\&\times
    \Theta(1-\Xp^2-z_1^2)
    \Theta(\Xp)
    \Theta\qty(q^2-q^2z_2^2-\qty(\frac{4+3q}{3+4q})^2\Xp^2)\notag\\
    &-
    \frac{1}{8}\frac{4+3q}{3+4q}
    \int\dd{z_1}
    \int\dd{z_2}
    \frac{\qty(\frac{4+3q}{3+4q})\Xp}{\qty(\frac{4+3q}{3+4q})^2\Xp^2+q^2z_2^2}
    \log(z_1^2)
    \delta\qty(\frac{z_1+qz_2}{1+q}-\Xeff)
    \notag\\&\times
    \Theta(\Xp)
    \Theta\qty(1-z_1^2)
    \Theta\qty(q^2-q^2z_2^2-\qty(\frac{4+3q}{3+4q})^2\Xp^2),
    \end{align}
where we use
\begin{equation}
    \int \frac{x}{x^2+a^2}\Theta(x_\mathrm{max}-x)\Theta(x-x_\mathrm{min})\dd{x}=
    \begin{dcases}
    \frac{1}{2}\qty[\log(x_\mathrm{max}^2+a^2)-\log(x_\mathrm{min}^2+a^2)]&x_\mathrm{max} > x_\mathrm{min}\\
    0&\mathrm{otherwise}.
    \end{dcases}
\end{equation}

Now we integrate either $z_1$ or $z_2$. Here, we integrate the variable that does not appear in the argument of the logarithmic function in each integral so that the content of the logarithmic function remains simple. Furthermore, we make the substitutions $z_1 = x$ and $z_2 = qx$ in each integral. Then, if we define
\begin{equation}
    F(x|a,b,c,d)\coloneqq\int_0^x\dd{x}\frac{b}{(x-a)^2+b^2}\log(\frac{x^2+c^2}{d^2}),
\end{equation}
one notices that all the integrals can be described with this function as
\begin{equation}
    \pi(\Xeff,\Xp|q) = I_1+I_2+I_3+I_4,
\end{equation}
where
    \begin{align}
    I_1 &= \frac{1+q}{8q}
    \Theta(\Xp)
    \notag\\&\times
    \int\dd{x}
    \pdv{}{x}\qty[F\qty(x\left|(1+q)\Xeff,\Xp,\frac{4+3q}{3+4q}\Xp,q\right.)]
    \Theta\qty(1-\Xp^2-(x-(1+q)\Xeff)^2)
    \Theta\qty(q^2-\qty(\frac{4+3q}{3+4q})^2\Xp^2-x^2)\\
    I_2 &= -
    \frac{1+q}{8q}
    \Theta(\Xp)
    \int\dd{x}
    \pdv{}{x}\qty[F\qty(x\left|(1+q)\Xeff,\Xp,0,q\right.)]
    \Theta\qty(1-\Xp^2-(x-(1+q)\Xeff)^2)
    \Theta\qty(q^2-x^2)\\
    I_3 &= \frac{1+q}{8q}\frac{4+3q}{3+4q}
    \Theta(\Xp)
    \notag\\&\times
    \int\dd{x}
    \pdv{}{x}\qty[F\qty(x\left|(1+q)\Xeff,\frac{4+3q}{3+4q}\Xp,\Xp,1\right.)]
    \Theta(1-\Xp^2-x^2)
    \Theta\qty(q^2-\qty(\frac{4+3q}{3+4q})^2\Xp^2-(x-(1+q)\Xeff)^2)\\
    I_4 &= -
    \frac{1+q}{8q}\frac{4+3q}{3+4q}
    \Theta(\Xp)
    \notag\\&\times
    \int\dd{x}
    \pdv{}{x}\qty[F\qty(x\left|(1+q)\Xeff,\frac{4+3q}{3+4q}\Xp,0,1\right.)]
    \Theta\qty(1-x^2)
    \Theta\qty(q^2-\qty(\frac{4+3q}{3+4q})^2\Xp^2-(x-(1+q)\Xeff)^2),
    \end{align}
After the tedious manipulations of step functions, we have \eqref{eq: I1} -- \eqref{eq: I4} and \eqref{eq: x1} -- \eqref{eq: x4}. So all we have to do is derive an analytical form of $F(x|a,b,c,d)$. It is obvious that $F(x|a,b=0,c,d)=0$. For $b\neq0$ case, The scaling $x=bt$ derives
\begin{equation}
    F(x|a,b,c,d)=\int_0^{x/b}\dd{t}\frac{\log(t^2+\frac{c^2}{b^2})-\log(\frac{d^2}{b^2})}{\qty(t-\frac{a}{b})^2+1}= \int_0^{x/b}\dd{t}\frac{\log(t^2+\frac{c^2}{b^2})}{\qty(t-\frac{a}{b})^2+1}+\log(\frac{b^2}{d^2})\qty[\arctan(\frac{x-a}{b})+\arctan(\frac{a}{b})].
\end{equation}
This means $F(x|a,b,c,d)$ can be reduced to 
\begin{equation}
    G(x|\alpha,\beta)\coloneqq\int_0^x\dd{x}\frac{\log(x^2+\beta^2)}{(x-\alpha)^2+1}.
\end{equation}
It can be shown that $G(x|\alpha,\beta)=-G(-x|-\alpha,\beta)$ so we can assume $x\geq0.$ This integral can be decomposed by a partial fractional decomposition of the complex range and a decomposition of the logarithmic function:
\begin{equation}
    G(x|\alpha,\beta)=\int_0^x\dd{x}\frac{\log(x^2+\beta^2)}{(x-\alpha)^2+1}
    =\mathrm{Im} \qty[\int_0^x\dd{x} \frac{\log(x-\beta i)}{x-\alpha-i}+\int_0^x\dd{x} \frac{\log(x+\beta i)}{x-\alpha-i}].
\end{equation}
We define these primitive functions as
\begin{equation}
    g(x|\alpha,\beta)\coloneqq\int^x\dd{t} \frac{\log(t-\beta i)}{t-\alpha-i}.
\end{equation}
By the transformation $t=\alpha+i-(\alpha+i-\beta i)s$,
\begin{equation}
    g(x|\alpha,\beta)=\int^{s=(\alpha+i-x)/(\alpha+i-\beta i)}\dd{s} \frac{\log[(\alpha+i-\beta i)(1-s)]}{s}.
\end{equation}
With the dilogarithm function
\begin{equation}
    \Dilog(z)=-\int_0^z\dd{z'}\frac{\log(1-z')}{z'},
\end{equation}
this integral has an analytical form
\begin{equation}
    g(x|\alpha,\beta)=\log(\alpha+i-\beta i)\log(\alpha-x+i)-\Dilog\qty(\frac{\alpha-x+i}{\alpha+i-\beta i})+C.
\end{equation}
Note that this expression is on the branch cut of complex logarithm function if we provide $\alpha\leq0$ and $\beta=1$. In addition, this expression may cross the branch cut of the dilogarithm function $\Dilog(z)$ at $z>1$. By examining the complex number that is the argument of the dilogarithm, we can see that this complex number takes a real number if $\beta\neq1$ and that real number is
\begin{equation}
    \left. \frac{\alpha-x+i}{\alpha+i-\beta i}\right|_{x=\frac{\alpha\beta}{\beta-1}}=\frac{1}{1-\beta}.
\end{equation}
We provide a threshold based on the relationship between $1/(1-\beta)$ and 1/2 so that the reflected dilogarithm argues complex number that passes through the real axis by 1/2 or less after the application of the reflection formula
\begin{equation}
    \Dilog(z)+\Dilog(1-z)=\frac{\pi^2}{6}-\log(z)\log(1-z).
\end{equation}
Finally, we have 
\begin{equation}
g(x|\alpha,\beta)
    =
    \begin{dcases}
        \log(x-\beta i)\log(\frac{\alpha -x+i}{\alpha +i- \beta i})+\Dilog\qty(\frac{x-\beta i}{\alpha +i-\beta i})& |\beta|<1\\
        \frac{1}{2}\qty(\log(x-\alpha-i))^2+\Dilog\qty(\frac{-\alpha}{x-\alpha-i})&\beta =1,\alpha\leq0\\
        \log(\alpha+i-\beta i)\log(\alpha-x+i)-\Dilog\qty(\frac{\alpha-x+i}{\alpha+i-\beta i})& \mathrm{otherwise}.\\
    \end{dcases}
\end{equation}
These expressions are identical up to terms containing $\alpha$ and $\beta$ but not $x$.
 
\end{widetext}

\bibliography{apssamp}%

\end{document}